\def\aap{\rm{A\&A}}

\documentclass[]{spie}

\usepackage{amsmath,amsfonts,amssymb}
\usepackage{array}
\usepackage{graphicx}
\usepackage[colorlinks=true, allcolors=blue]{hyperref}
\usepackage{placeins}

\newcolumntype{L}[1]{>{\raggedright\arraybackslash}p{#1}}
\hypersetup{
  pdftitle={STarlight Acquisition and Reflection toward Interferometry (STARI) mission},
  pdfauthor={John D. Monnier et al.},
  pdfsubject={Author version of Proc. SPIE 14145, 1414512 (2026)},
  pdfkeywords={space interferometry, CubeSat, formation flying, single-mode fiber}
}

\title{STarlight Acquisition and Reflection toward Interferometry (STARI) mission}

\author[a]{John D. Monnier}
\author[a]{James Cutler}
\author[a]{Akshat Dubey}
\author[a]{Caden Burkhardt}
\author[a]{Raven St. Clair}
\author[a]{Bradley Bialke}
\author[a]{Adrien Boissy}
\author[a]{Tobey Chang}
\author[a]{Hannah Cherry}
\author[a]{Yethinder Ragav Lakshmi Kumar}
\author[a]{Michael Meyer}
\author[a]{Jacob Klingler}
\author[a]{Shivani Sunil}
\author[a]{Austin Thomas}
\author[a]{Daniil Voloshin}
\author[b]{Simone D'Amico}
\author[b]{Antonio Rizza}
\author[b]{Ethan Foss}
\author[b]{Samuel Y.W. Low}
\author[b]{Prachet Jain}
\author[c]{E. Glenn Lightsey}
\author[c]{Althea Noonan}
\author[c]{Mark Gorrow}
\author[d]{Leonid Pogorelyuk}
\author[d]{Melissa Flores}
\author[d]{Albert Ostoja-Starzewski}
\author[d]{Garret DeRose}
\author[d]{Jasper Heymann}
\author[d]{Alexa LaPinta}
\author[d]{Ryan Turoff}
\author[e]{Gautam Vasisht}
\author[f]{Kerri Cahoy}
\affil[a]{University of Michigan, Ann Arbor, MI, USA}
\affil[b]{Stanford University, Stanford, CA, USA}
\affil[c]{Georgia Institute of Technology, Atlanta, GA, USA}
\affil[d]{Rensselaer Polytechnic Institute, Troy, NY, USA}
\affil[e]{NASA Jet Propulsion Laboratory, Pasadena, CA, USA}
\affil[f]{Massachusetts Institute of Technology, Cambridge, MA, USA}

\authorinfo{Corresponding author: John D. Monnier, \href{mailto:monnier@umich.edu}{monnier@umich.edu}.
Author version: \href{https://doi.org/10.1117/12.3104103}{Proc. SPIE \textbf{14145}, 1414512 (2026)}.}
\begin{document}
\maketitle
\thispagestyle{plain}

\begin{abstract}
We introduce the NASA-funded mission STARI (STarlight Acquisition and Reflection toward Interferometry), a two-spacecraft technology demonstrator for future long-baseline interferometers. STARI will perform, to our knowledge, the first in-space transfer of starlight between free-flying spacecraft, a critical step toward long-baseline space interferometry. Two propulsive 8U CubeSats in low Earth orbit will fly in formation with separations up to 200 m: one relays a narrow stellar beam to the other, where an off-axis parabola reimages it into a single-mode fiber. Using differential GNSS, visible LED beacons, and fast-steering mirrors in closed-loop control, we aim to sustain diffraction-limited, high-throughput fiber coupling during formation flight. STARI, targeting launch in 2029, is designed to mature key technologies needed for
future mission concepts such as the Large Interferometer for Exoplanets
(LIFE), which would seek atmospheric biomarkers on nearby Earth-like
exoplanets.
\end{abstract}


\begin{figure}[t]
\centering
\includegraphics[width=0.95\linewidth]{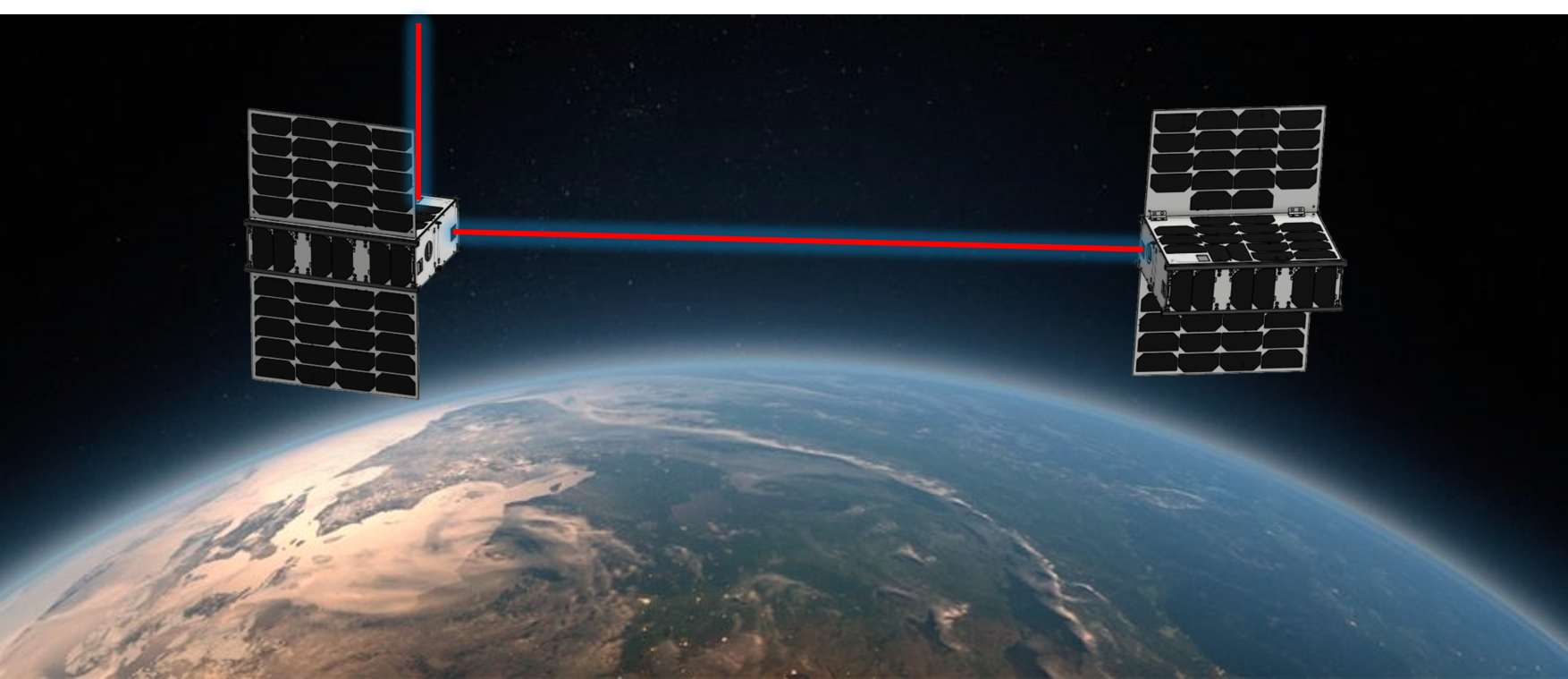}
\caption{Current PDR rendering of the STARI two-spacecraft handoff.
One spacecraft, acting as Pupil Transport (PT), collects starlight and
reflects a narrow beam across the formation to the partner spacecraft,
acting as Fiber Injection and Detection (FID), where the beam is captured
by the optical train and coupled into a single-mode fiber.}
\label{fig:mission_overview}
\end{figure}

\section{Introduction}
\label{sec:intro}

Long-baseline interferometry bypasses the diffraction limit of a single filled aperture by coherently combining light collected across a much larger physical baseline. Ground-based interferometers have demonstrated milli-arcsecond imaging of stellar surfaces and circumstellar environments \cite{monnierreview2003}, but atmosphere, weather, and geography limit both sensitivity and baseline flexibility. Space interferometry removes these barriers and is especially compelling for future mid-infrared nulling missions that could measure atmospheric biomarkers on nearby terrestrial exoplanets \cite{tpfi2008,alei2022,Quanz2022LIFEI}. Recent formation-flying interferometry roadmaps and exoplanet-interferometry study reports identify flight validation of beam transfer, optical coupling, path-length stability, and coordinated spacecraft control as a key step toward such observatories \cite{monnier2019_whitepaper_ff,quanz2026_kiss_interferometry}. The remaining challenge is not conceptual, but practical: future space interferometers require precision formation control, beam transfer, optical coupling, autonomous operations, and flight-proven performance data at the system level.

STARI is a pathfinder mission built around one essential operation for a future distributed optical interferometer: collect starlight on one free-flying spacecraft, transport that beam across tens to hundreds of meters, and inject it into a single-mode fiber on another spacecraft. Figure~\ref{fig:mission_overview} illustrates the present handoff concept. The mission is not itself an interferometer. It deliberately omits a  combiner and delay line so that the project can concentrate on the flight demonstrations that are difficult to reproduce on the ground: relative navigation in low Earth orbit, optical beam handoff, closed-loop fine steering, coupling stability, station-keeping using propulsion maneuvers, low platform jitter, and autonomous coordinated two-spacecraft operations.

A 2024 SPIE paper described STARI during its pre-Phase-A CubeSat stage \cite{stari2024} before being officially selected for funding, with a start date of December 2024.  A 2025 SmallSat/SRR update captured the first NASA-funded baseline: a 6U-class design, a $\sim 100$ m starlight-transfer demonstration, a reference-laser-assisted acquisition concept, and a maturing ConOps for transitions between standby and science relative orbits. Since then, the mission has carried out the System Requirements Review (SRR) in Ann Arbor in December 2025 and passed after responding to action items identified by the review committee; the project then carried out an internal preliminary design review (PDR) in May 2026 on the campus of Rensselaer Polytechnic Institute in Troy, New York. Here in the 2026 SPIE proceedings, we summarize the present design and the main changes since 2024, with emphasis on how the subsystem trades are converging toward a flight experiment.

\begin{table}[ht]
\caption{STARI at a glance. Values are approximate working values from the
May 2026 preliminary-design baseline. The table summarizes the mission scale
and current design assumptions rather than the full requirements traceability
matrix.}
\label{tab:basic_params}
\centering
\begin{tabular}{|L{1.85in}|L{4.05in}|}
\hline
Parameter & Current working value \\
\hline
Spacecraft & Two functionally redundant propulsive 8U CubeSats \\
\hline
Mass & Approximately 12 kg current estimate per spacecraft \\
\hline
Orbit & 500--600 km low Earth orbit, with a near-terminator or dawn-dusk SSO preferred \\
\hline
Mission duration & At least 6 months, with extended operations possible if spacecraft health and fuel permit \\
\hline
Formation scale & Nominal starlight handoff between $20$--$100$ m separations; extended demonstrations $>175$ m \\
\hline
Payload details & 2U volume, with a 20-mm beam aperture with fast-steering mirrors and single-mode-fiber injection \\
\hline
Propulsion & R236fa cold-gas system; mission $\Delta v$ budget is in the tens of m/s class and remains coupled to GNC and packaging trades \\
\hline
Navigation & Differential GNSS, LED-beacon imaging, payload camera centroids, inter-satellite link \\
\hline
\end{tabular}
\end{table}

\section{Mission concept at a glance}
\label{sec:mission_glance}

The basic observing task is to acquire a target star, send the beam across
the formation, search for single-mode-fiber coupling, and record the
coupled flux to characterize the overall throughput and stability. In
addition, the GNC system must maintain an ``interferometer'' geometry
where the baseline between spacecraft is perpendicular to the line of
sight within the dynamic environment of low Earth orbit. While not a full
interferometer, STARI is a flight demonstration of a single interferometer
arm: free-space optical handoff between formation-flying spacecraft,
coupling into single-mode fiber, and cm-level formation-flying control.
These are key technology gaps identified in the recent KISS workshop report
\cite{quanz2026_kiss_interferometry} {\em Exploring Exoplanets with
Interferometry}.

Table~\ref{tab:basic_params} summarizes the current working mission
configuration and gives the scale of the experiment. It is not intended
to reproduce the full requirements traceability matrix, but to provide
the spacecraft, orbit, payload, formation, propulsion, navigation, and
operations context for the design choices discussed below.

The mission-success criteria are staged to reflect increasing levels of
technology maturation and system-level validation. Threshold success demonstrates the essential building
blocks: repeated starlight handoff between free-flying spacecraft,
autonomous transfer between passively safe standby and science
formations, and in-flight characterization of propulsion, pointing
jitter, and payload vibration. Baseline success adds sustained
performance during science operations, including OPD control below
50 cm, stable handoff over multiple baselines, fiber-coupled starlight
from $V \simeq 2.5$ mag targets with $\leq 40\%$ rms coupling variation
over 1 s windows, and 10-mm-class lateral relative-orbit knowledge.
Extended goals push the architecture toward future interferometer
needs by attempting handoff and fiber injection beyond 175 m,
cm-class active formation control with sub-mm knowledge, and rapid
retargeting between nearby stars.

\section{Evolution since the 2024 concept}
\label{sec:evolution}

STARI is now planned around two functionally redundant 8U spacecraft, with a 2U optical payload and a 1.6U propulsion system.  Replacing the back-propagated laser reference beacons, we  will use visible LED beacons for relative optical sensing by a fast-readout camera coupled with a fast-steering mirror to compensate for pointing jitter and drifts in relative orbit.  We will still use the direct measurement of fiber-coupled starlight as the
key mission observable, though the present design uses a camera for detection
rather than the avalanche-photodiode option considered in the earlier STARI
concept \cite{stari2024}, an approach with relevant CubeSat heritage
from PicSat \cite{picsat}.

The 2026 design also clarifies terminology. The spacecraft acting as Pupil Transport (PT) collects starlight from the target star and relays a narrow beam across the formation. The spacecraft acting as Fiber Injection and Detection (FID) receives the relayed beam and injects it into a single-mode fiber.
These optical roles are independent of the formation-flying roles: the
passive spacecraft provides the inertial reference, while the active
spacecraft performs the propulsive maneuvers needed to adjust the relative
orbit. Either spacecraft is intended to be capable of taking either optical
role, PT or FID, or either formation-flying role, active or passive, based
on fuel state, subsystem health, or operations needs.
 This symmetry increases development complexity but decreases single-failure risk and reduces non-recurring engineering by using a common spacecraft design.

\section{Collaboration and subsystem contributions}
\label{sec:team}

The STARI team is organized around subsystem responsibilities and institutional expertise. The University of Michigan leads mission and systems engineering, the Pupil Transport (PT) payload, payload avionics, spacecraft-bus interfaces, and mission ground control operations. Stanford University leads formation flying and GNC:
calculation and maintenance of standby and science relative orbits,
``zero-optical-path-difference'' interferometry geometry,
differential GNSS and beacon-aided navigation, maneuver planning,
and safety logic. These Stanford-led developments are described in
companion work on STARI formation design using relative orbital
elements \cite{Rizza2026STARIFormationDesign}, flight-ready
carrier-phase GNSS navigation for distributed space systems
\cite{Low2026FlightReadyGNSS}, and input-constrained impulsive
control for spacecraft relative motion
\cite{Foss2026ImpulsiveOptimalControl}.
Georgia Tech leads the cold-gas propulsion system, adapting the VISORS
3-D-printed cold-gas propulsion heritage \cite{Hart2024VISORSPropulsion}
to STARI's $\Delta v$, minimum-impulse, thermal, and packaging constraints;
the broader VISORS mission also provides useful formation-flying and autonomous
GNC heritage \cite{Guffanti2023VISORSGNC}.
 Rensselaer Polytechnic Institute leads the fiber injection and detection (FID) optical subsystem, including the off-axis parabola, focus mechanism, and single-mode fiber injection. JPL and MIT provide collaborator review and expertise in interferometry, small-spacecraft systems, payload operations, and mission design.

\section{STARI hardware architecture}
\label{sec:architecture}

\noindent{\bf Spacecraft bus:} STARI is being designed as a pair of nearly identical 8U CubeSats. Each spacecraft carries the bus functions needed for power, command and data handling, attitude determination and control, communications, thermal management, and structure. Each also carries the optical payload elements needed to act as either PT or FID. That symmetry keeps the two vehicles common, but it also means the payload, propulsion, thermal, and software interfaces must be packaged twice in a very small volume.  We are currently in the process of finalizing the selection of a commercial partner to provide the STARI buses.

\begin{figure}[ht]
\centering
\includegraphics[width=0.9\linewidth]{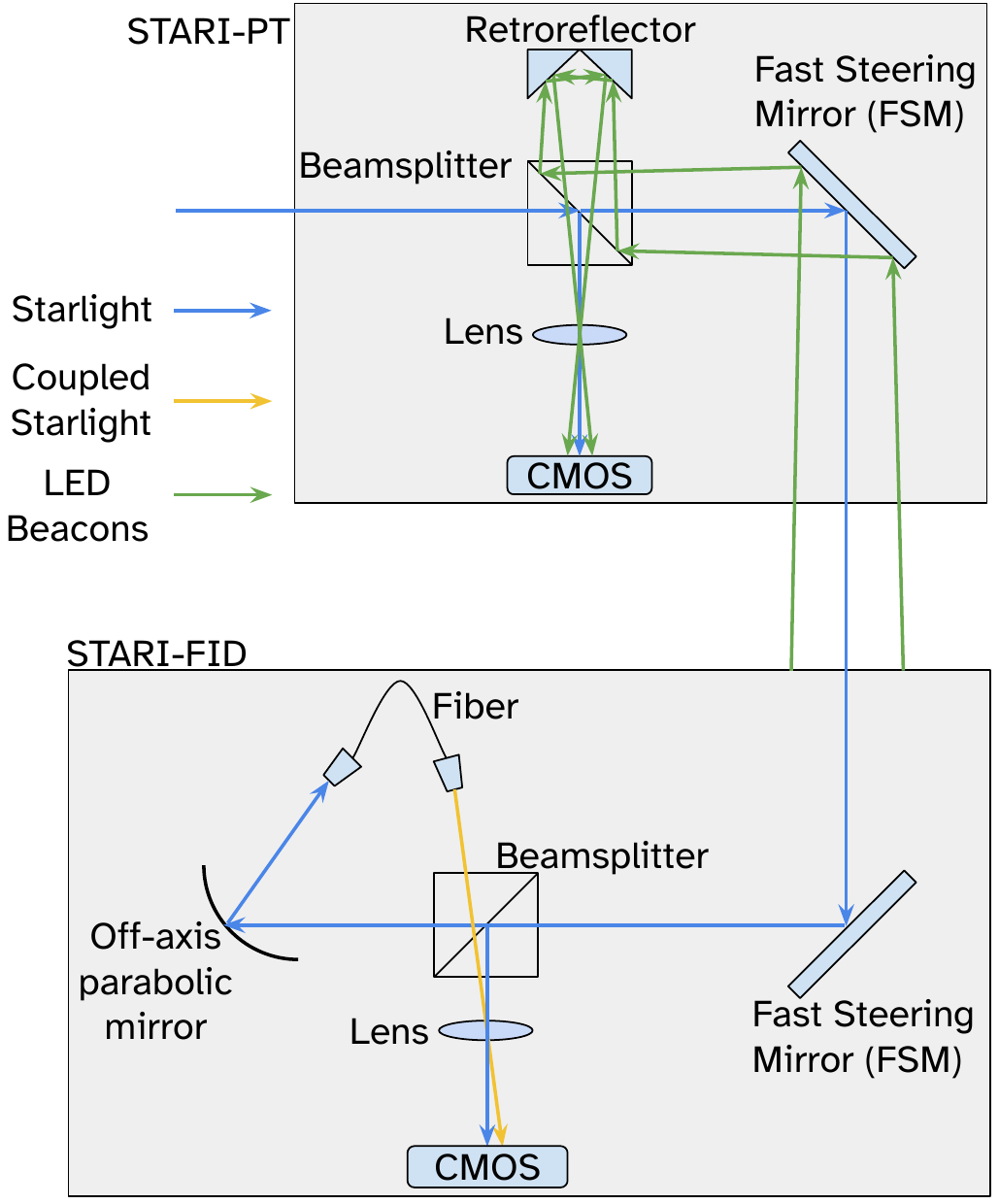}
\caption{The optical handoff as a block diagram. PT sees the star and the partner-spacecraft LED beacons in the same camera frame, then uses a fast-steering mirror to send the beam toward FID. FID receives that beam, steers it locally, injects it into a single-mode fiber with an off-axis parabola, and images the fiber-coupled light so the mission can measure coupled-flux stability directly.}
\label{fig:payload_block}
\end{figure}

\noindent{\bf Payload architecture:} A more detailed functional diagram of the payload-level architecture is shown schematically in Figure~\ref{fig:payload_block}. In this architecture,
PT uses a beamsplitter and retroreflector to image the target star and
the FID optical beacons in the same camera frame, allowing the payload loop
to correct for attitude jitter and uncertainty in the physical location of
FID. PT then steers the outgoing beam with a fast-steering mirror, and FID
receives the relayed beam.
 The FID then controls a second fast-steering mirror to inject the starlight into a single-mode fiber. The coupled light is re-imaged on a detector so that the mission can measure coupling stability directly, not merely infer alignment from centroid positions. The wavelength split that supports these sensing and fiber-injection functions is discussed in Section 7.

\begin{figure}[ht]
\centering
\includegraphics[width=\linewidth]{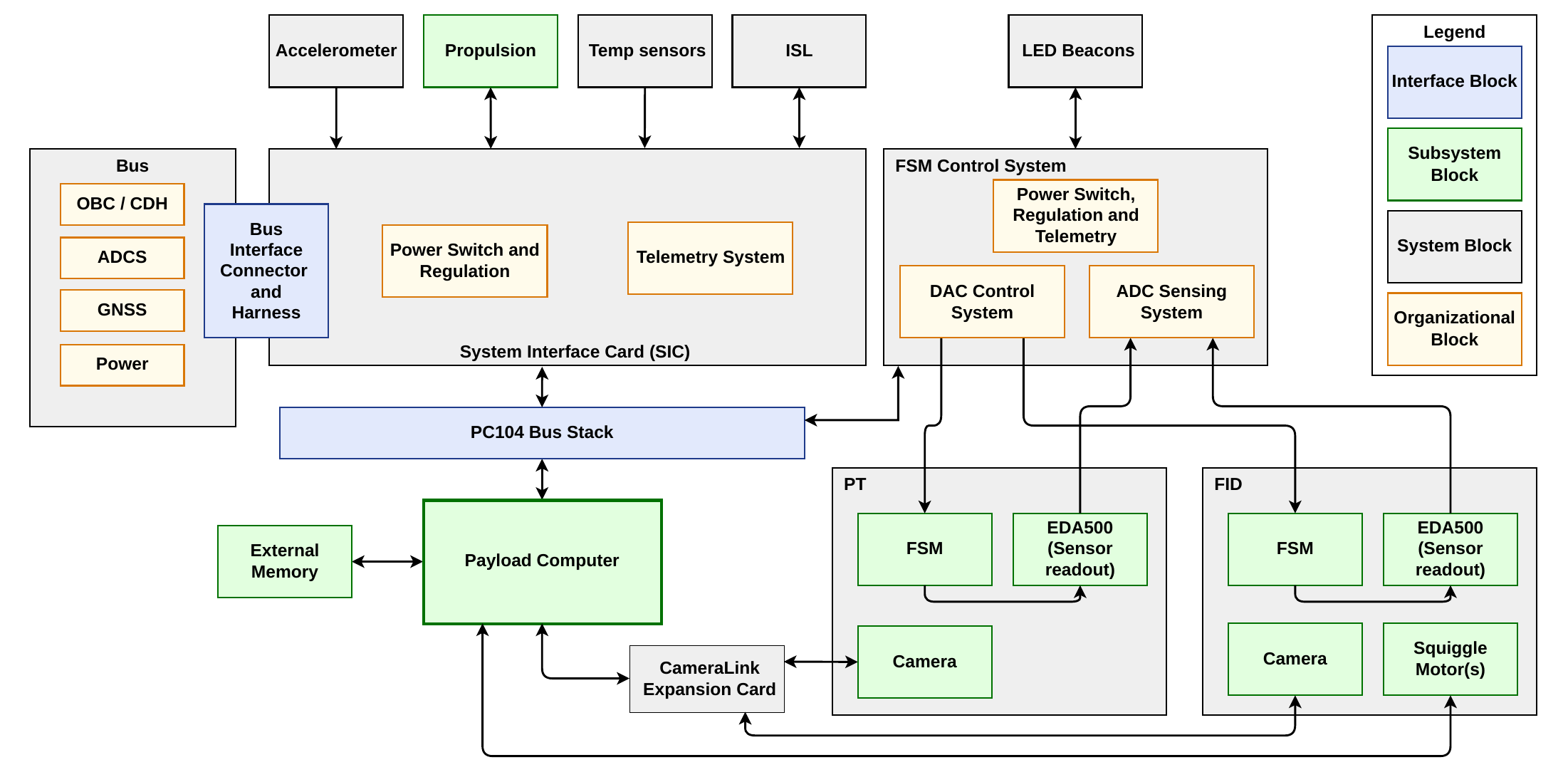}
\caption{
Simplified architecture for STARI payload avionics and control.
The diagram omits individual harness and connector details, but shows the
main power, data, command, sensing, and diagnostic paths among the bus
interface, payload computer, cameras, FSM control system, LED beacons,
accelerometers, inter-satellite link, PT/FID subsystems, and propulsion
interface.
}

\label{fig:payload_avionics}
\end{figure}

\noindent{\bf Payload avionics and interfaces:} The payload is not a passive passenger on the spacecraft bus. While the bus computer handles conventional spacecraft functions like power generation/storage, ground communications, and star tracking, the payload computer hosts camera acquisition, FSM control, LED beacon control, intersatellite communications, payload telemetry, GNC processing, and propulsion control.
The full connector and interface-control diagram is too detailed to reproduce here, since it includes individual power, timing, command, telemetry, and diagnostic connections among the bus computer, payload computer, cameras, fast-steering mirrors, beacon drivers, inter-satellite link, GNC software, and propulsion interface. Figure~\ref{fig:payload_avionics} therefore shows a simplified functional architecture, emphasizing the signals that must close during an observation: camera and beacon measurements, GNC/FSM commands, and payload telemetry.

\section{Concept of operations}
\label{sec:conops}

\noindent{\bf Orbit selection:} STARI is designed for low Earth orbit, with a dawn-dusk or near-terminator sun-synchronous orbit (SSO) between about 500 and 600 km preferred. That orbit is attractive because it balances power generation, target viewing geometry, drag, deorbit compliance, and rideshare availability. That said, the mission does not require any particular local time of ascending node (LTAN) orbit  and we will be able to meet all requirements in any SSO.  The target list and observing cadence can be adapted once the actual launch opportunity and local time of ascending node are known.

\begin{figure}[ht]
\centering
\includegraphics[width=.48\linewidth]{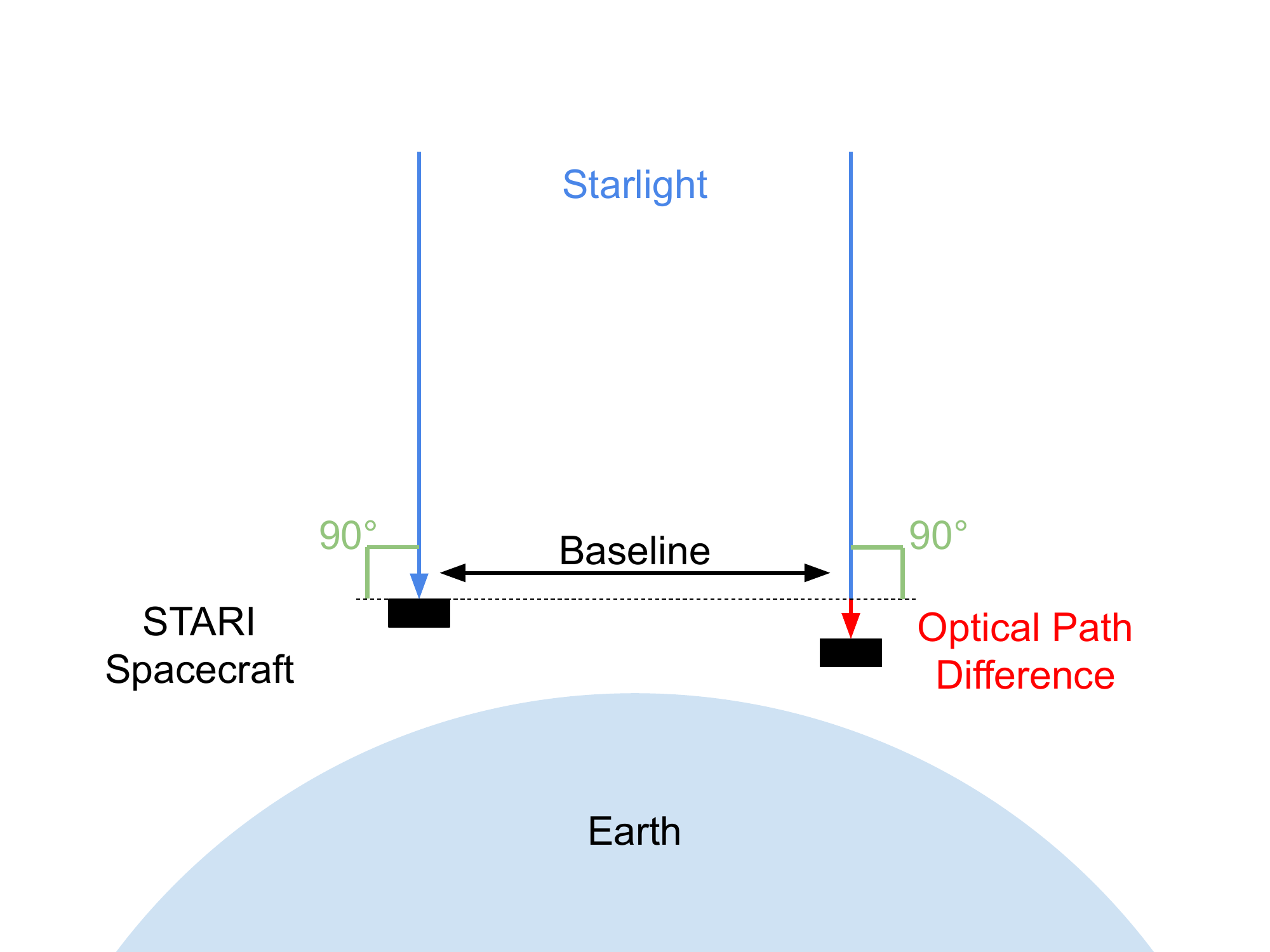}
\includegraphics[width=0.48\linewidth]{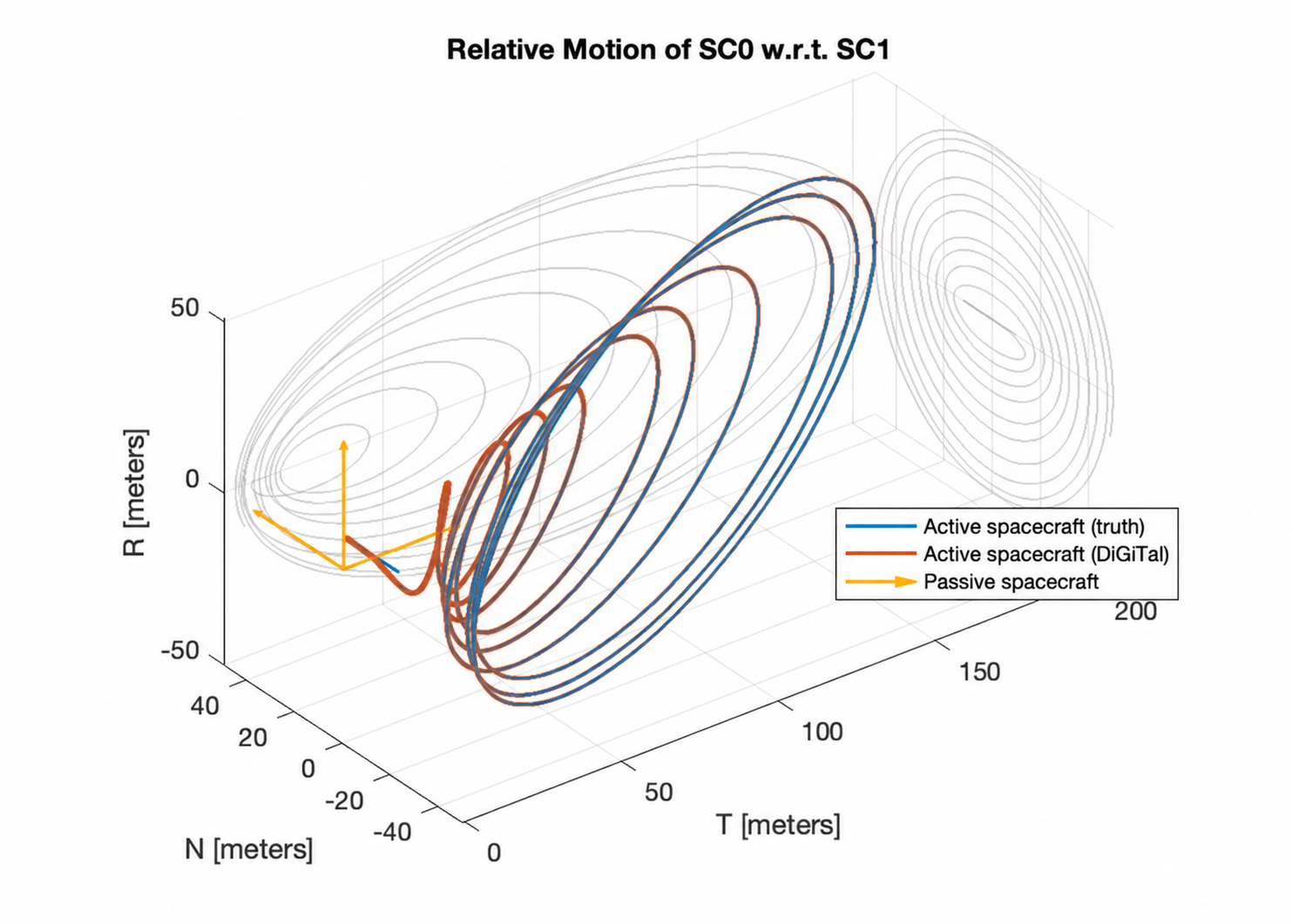}
\caption{(a) Definition of optical path difference (OPD) and baseline.
For an interferometer, we want the spacecraft to stay within a plane
perpendicular to the star's direction, giving an OPD near 0. This allows
a future space interferometer to fly with only compact differential delay
lines, a huge advantage over ground-based facilities. (b) 3D relative
trajectory in RTN during transfer from science to standby \cite{Rizza2026STARIFormationDesign}. The legend
compares the simulated truth trajectory with the DiGiTal navigation
solution, where DiGiTal denotes Distributed Multi-GNSS Timing and
Localization. The plot is also a reminder that science geometry, maneuver
timing, fuel cost, and collision safety are one coupled trade.}
\label{fig:opd}
\end{figure}

\noindent{\bf Orbit design:} The Stanford-led formation-flying strategy is driven by minimizing the
optical path difference (OPD) of the observing geometry; the detailed
relative-orbit design is presented by Rizza et al.
\cite{Rizza2026STARIFormationDesign}.
 The definition of OPD can be seen in Figure\,\ref{fig:opd}a, where an OPD of 0\,m occurs when the vector separating the spacecraft is exactly 90 degrees from the star's direction.  In a future interferometer, a delay line need only compensate finite OPD mismatch and so the OPD does not need to be truly zero. STARI therefore uses the relative orbit to keep the two spacecraft close to a plane perpendicular to the line of sight to the star. Figure~\ref{fig:opd}b shows an example relative motion design from the Stanford GNC analysis. The design space includes passively safe standby orbits, transfers between standby and science configurations, science station keeping, and reactive safety. Differential GNSS provides the baseline navigation layer, building on
carrier-phase GNSS navigation software developed for distributed space
systems \cite{Low2026FlightReadyGNSS}, while the LED beacons and
payload camera centroids provide the lateral information needed for
higher-precision optical alignment and extended goals.

\noindent{\bf Observing sequence:} A typical observing attempt begins in a passively safe standby configuration. From there, the spacecraft transfer into a science relative orbit, where the baseline is chosen to lie nearly perpendicular to the target-star direction in order to keep the OPD small. After each science experiment (typically 2-3 orbits in duration), the pair returns to standby. Safe and escape modes provide fallback paths if navigation, communications, attitude, propulsion, or payload state no longer matches the planned sequence.  Generally, these maneuvers will be programmed ahead of time and executed autonomously, with active monitoring of telemetry mainly during the science orbit phase where the formation is not passively safe and requires active control of the formation.

\begin{figure}[hb]
\centering
\includegraphics[width=0.45\linewidth]{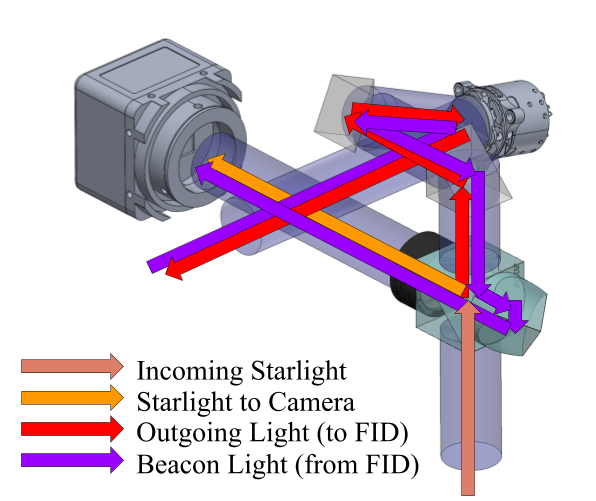}
\includegraphics[width=0.45\linewidth]{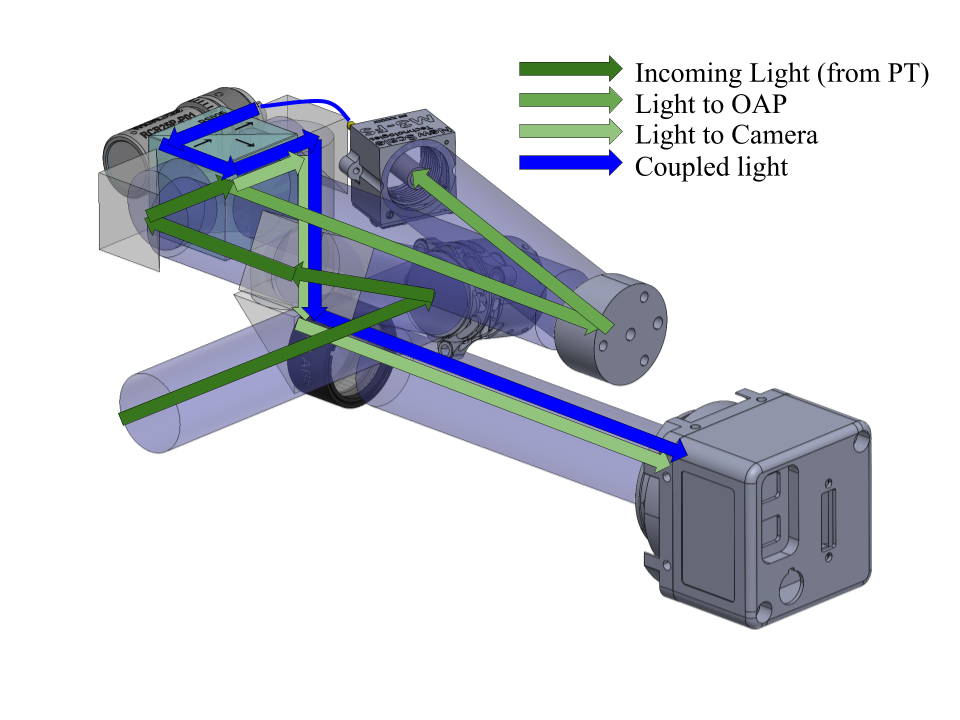}
\caption{These open-frame CAD views emphasize the PT optics with beam path of the star on the left (coming from the bottom), and FID optics with path of the reflected starlight from the PT on the right. Mounting plates and structures are omitted for clarity.}
\label{fig:pt_cad}
\end{figure}

\section{Payload subsystem status}
\label{sec:payload}

\noindent{\bf Optomechanical design of payload:} We have already described the overall design and functionality of the PT and FID payloads.  Figure~\ref{fig:pt_cad} shows the current CAD design of how we fit the full functionality of Figure~\ref{fig:payload_block} into a 2U volume. The model contains two cameras, two fast-steering mirrors, along with a fiber-focusing module using an off-axis parabola.  Although the mounting plates are omitted for clarity, the beam paths have
been routed to avoid CubeSat structural elements.  We are fabricating plates for mounting the hardware in summer 2026
for testing and characterization in preparation for our Design Review in 2027.

\noindent{\bf Candidate payload hardware:} Implementation studies are converging on a small set of reference components: FSO FSM20D-class fast-steering mirrors, Raptor Hawk 1920-class low-noise cameras, and a payload computer architecture based on a CAVU OBC Polar with CameraLink expansion board and dedicated FSM interface electronics.

For FID, the athermal fiber-injection module will likely use either a
custom Zerodur OAP mounted on Invar brackets or an on-axis achromatic
doublet.
 We plan to include a focus mechanism so that thermal or launch-induced shifts can be recovered by an on-orbit search and calibration procedure.

\begin{figure}[h]
\centering
\includegraphics[width=0.6\linewidth]{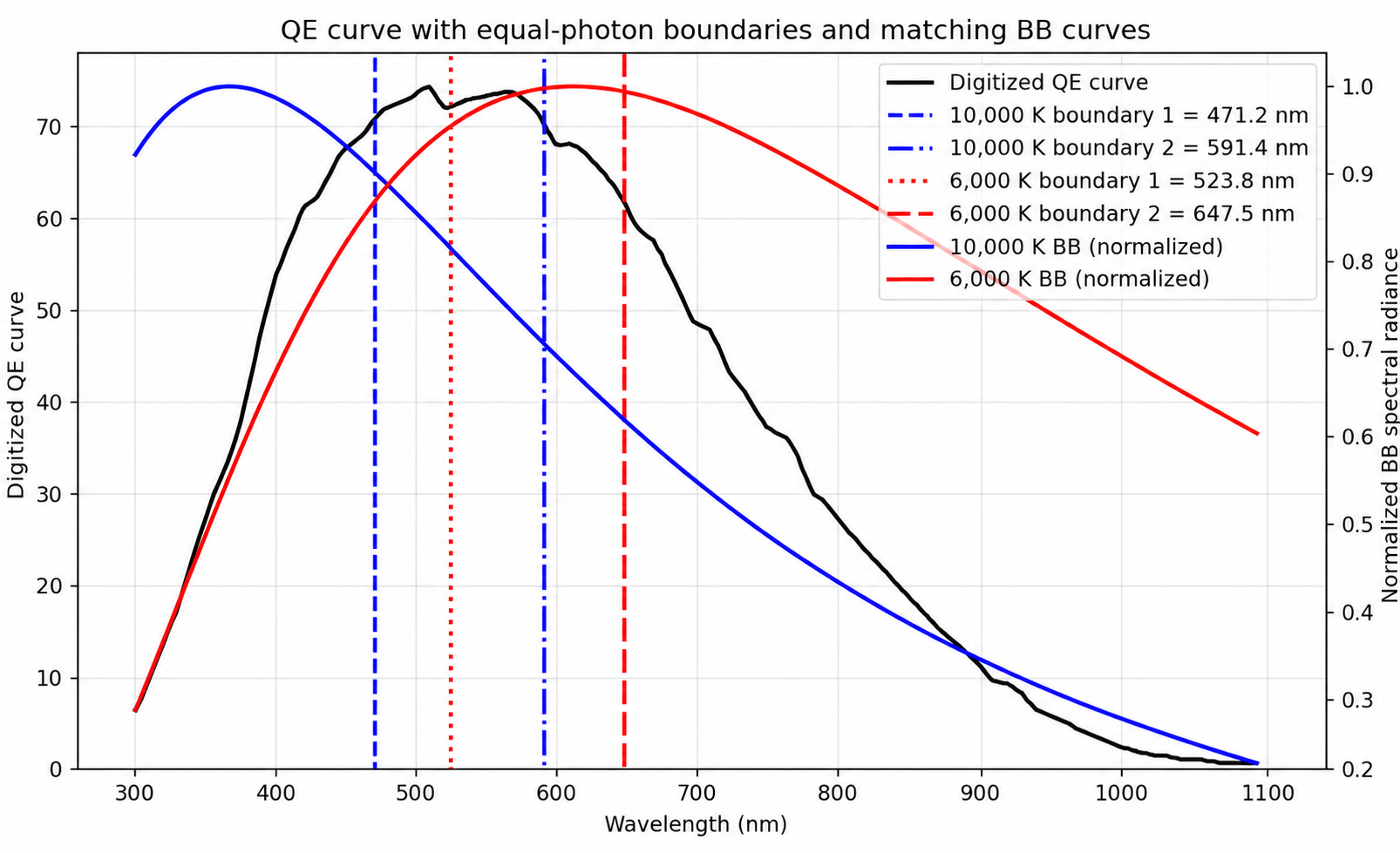}
\caption{Payload spectral split used for photon-budget trades. The plot
combines representative stellar blackbody (BB) spectra with detector quantum efficiency to
estimate detected photon rates in the PT sensing, FID tracking loop, and finally the fiber-injection coupling measurement. The vertical boundaries show example equal-photon
splits for representative stellar temperatures, illustrating that a
blue/green/red division of the visible band can provide comparable photon
rates in the three payload channels for bright targets. Final dichroic
edges, coating losses, and fiber throughput remain active trades.}
\label{fig:wavelength_snr}
\end{figure}

\noindent{\bf Wavelength split and photon budget:} The payload uses the visible band in three pieces. Blue light below roughly 500 nm feeds PT acquisition and beacon/star centroiding; a green visible band near 500--630/650 nm is reserved for the single-mode fiber; and redder light supports FID tip/tilt sensing. The exact dichroic and fiber edges remain a trade, but the calculation in Figure~\ref{fig:wavelength_snr} shows why this split is plausible. For a $V=2.5$ mag star and a 20 mm aperture, the equal-photon calculation gives roughly $3\times10^5$ detected photons s$^{-1}$ in each of the three color partitions. In the 6000 K case, boundaries near 524 and 648 nm give about $2.95\times10^5$ photons s$^{-1}$ in the blue, green, and red channels; for a 10,000 K star, the boundaries shift to about 471 and 591 nm with about $3.1\times10^5$ detected photons s$^{-1}$ per channel. For 20 ms exposures, the PT blue-band estimate is therefore about 6000 photons per frame, adequate for arcsecond-precision closed-loop tracking for our target limit  of $V = 2.5$ mag.

\begin{figure}[!htbp]
\centering
\includegraphics[width=0.95\linewidth]{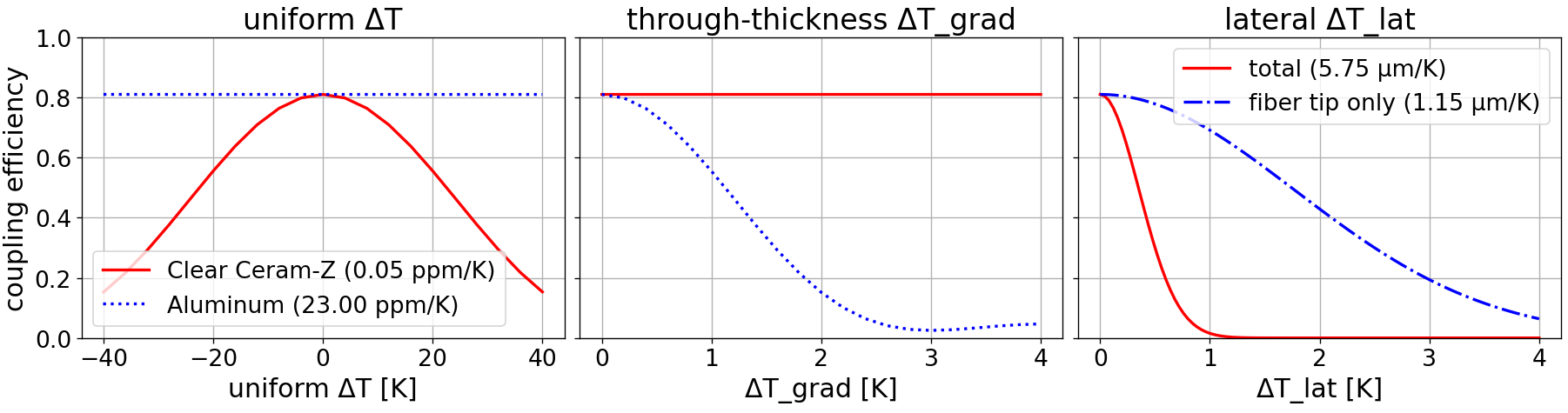}
\caption{Sensitivity of the FID fiber-coupling efficiency to thermally
induced focus and alignment errors in the OAP/fiber module. The curves
illustrate that small focus shifts, lateral offsets, or thermal gradients
can rapidly reduce coupled flux. This sensitivity motivates the FID focus
mechanism, careful material choices, thermal modeling, and an on-orbit
coupling search/calibration procedure.}
\label{fig:fid_thermal}
\end{figure}

\noindent{\bf Payload trades:} Fiber coupling is sensitive to tip-tilt, focus, lateral fiber placement, thermal gradients, and launch or thermal distortion of the OAP/fiber assembly. The coupling-efficiency curves in Figure~\ref{fig:fid_thermal} show why thermal and material choices matter: small focus or lateral shifts can quickly become a coupled-flux loss, motivating an active focus/search procedure and thermal modeling. PT simulations using realistic jitter spectra indicate promising sub-arcsecond residual tracking in simplified cases, but the integrated model still has to account for camera noise, defocus, optical transmission, spacecraft jitter, relative-orbit drift, and true payload-control latencies.

\paragraph{Optical-component vibration diagnostics:}
Although STARI will validate free-space beam handoff and fiber injection, it will not measure interferometric phase directly. For future space interferometers, however, internal motion of payload optics can be as important as external formation control: sub-micron optical-path changes correspond to non-negligible phase errors in the visible and infrared. We are therefore studying a secondary technology objective in which triaxial accelerometers would be mounted on selected optical components or optical benches to measure component-level vibration during normal operations. Candidate disturbance sources include reaction-wheel jitter, cold-gas thruster impulses, solar-array motion, and propellant slosh. Tests of CubeSat-class reaction wheels have shown vibration spectra with significant content out to 500 Hz, motivating low-noise spectral measurements with piezoelectric or MEMS accelerometers \cite{Shields2017CubeSatReactionWheelAssemblies}. Lower-frequency and quasi-static disturbances from propulsion events are more naturally captured in the time domain, for which MEMS accelerometers are attractive. The final sensor complement remains under trade, but preliminary analyses indicate that such diagnostic measurements are feasible and could provide useful vibration data for future free-flying interferometer designs.

\section{Bus-related subsystem status}
\label{sec:subsystems}

\paragraph{Propulsion:}
The Georgia Tech-led propulsion subsystem must provide enough total impulse
to reconfigure the formation while also delivering repeatable small impulses
for precision station keeping\cite{noonan2026design}.

The current concept is a cold-gas system with main and plenum tanks, valve manifold, nozzles, heaters, pressure/temperature telemetry, and a controller board. It builds on Georgia Tech VISORS heritage while adapting the design to the STARI bus, $\Delta v$, minimum-impulse, and packaging constraints. The working propellant choice is R236fa, and the near-term test campaign focuses on refill timing, minimum impulse, repeatability, thermal behavior, leak performance, and thrust-stand characterization. Two thermal-feed architectures are being tested in 2026 before the project selects the flight architecture for the January 2027 CDR and 2027 build phase:
\begin{itemize}
\item Insert a heater into the plenum so the gas is preheated before the nozzle valve opens.
\item Add heaters at the inlets of valves plumbed directly to the gas reservoir, removing the plenum from that branch.
\end{itemize}

\noindent{\bf GNC/propulsion interface:} Propulsion is a good example of why STARI is a coupled systems-engineering problem. Canted nozzle placement can create cosine losses that increase $\Delta v$ demand, while placement far from the spacecraft center of mass can introduce parasitic torques and reaction-wheel momentum concerns.
The minimum impulse bit, currently tracked near 2 mN\,s or
0.166 mm\,s$^{-1}$ for a 12 kg spacecraft, matters for the most
ambitious formation-control goals and feeds directly into the
input-constrained maneuver-planning problem
\cite{Foss2026ImpulsiveOptimalControl}.

\begin{figure}[t]
\centering
\includegraphics[width=0.72\linewidth]{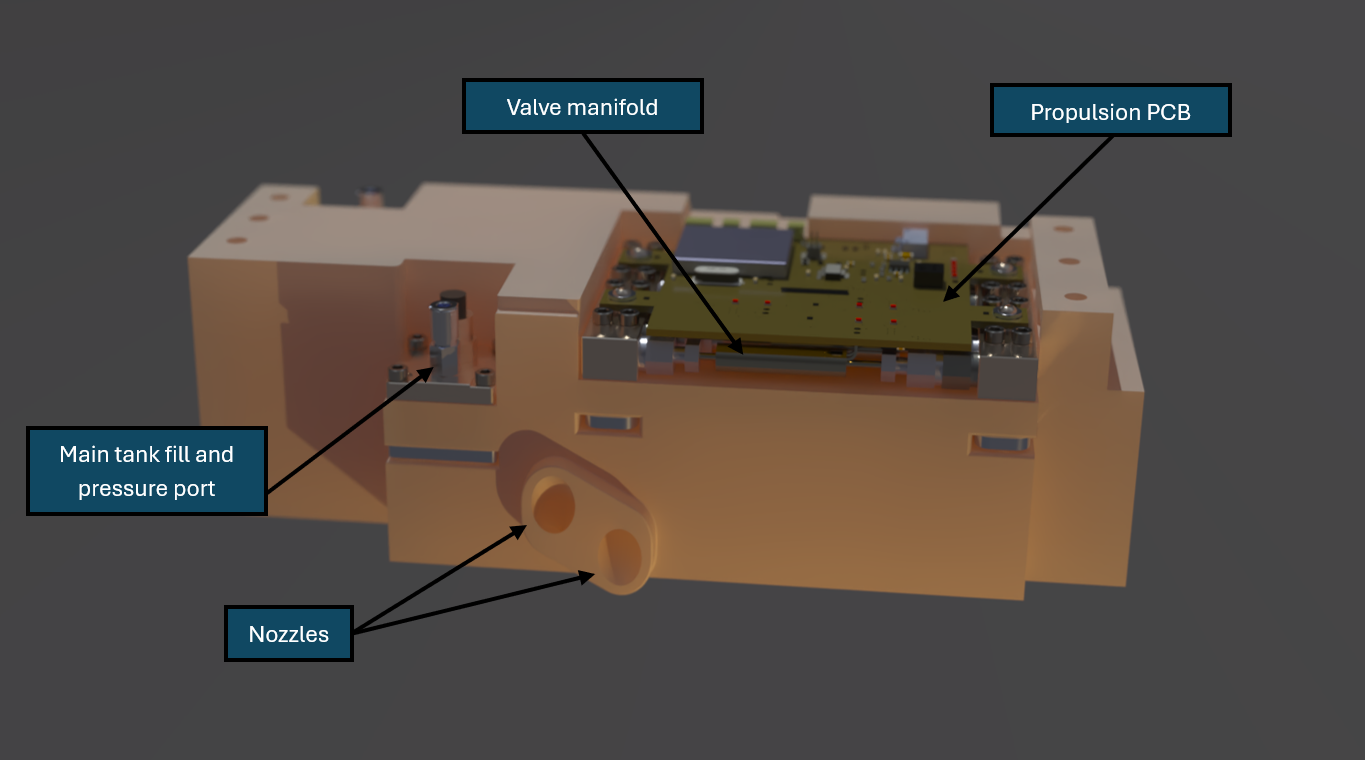}
\caption{Annotated rendering of the STARI cold-gas propulsion prototype
concept. The model highlights the tank fill/pressure port, valve manifold,
nozzles, and propulsion control electronics. Because propulsion impulses set
the achievable formation-control precision, nozzle placement, impulse
repeatability, thermal behavior, and refill logic remain coupled to the GNC
and payload-control trades.}
\label{fig:propulsion}
\end{figure}

\noindent{\bf Avionics and communications:} The avionics and communications design has matured around the split between ordinary spacecraft housekeeping and payload control. The bus computer remains responsible for conventional spacecraft control and safe-mode functions, while the payload computer handles camera acquisition, FSM control, LED beacons, payload telemetry, and interfaces to GNC. Communications include telemetry, tracking, control, high-speed downlink, and an inter-satellite link (ISL). The ISL is especially important: relative navigation, payload state, planned maneuvers, and fault notifications have to move between spacecraft fast enough to support autonomous formation operations.

\section{Performance simulations}
\label{sec:verification}

\noindent{\bf Performance simulations:} The goal of our simulation effort is to combine attitude control, relative orbit control, line-of-sight sensing, payload-camera centroiding, FSM response, optical throughput, and fiber coupling. The output of interest is the coupled-flux time series: how much light gets into the fiber, how stable it is, and which spacecraft or payload disturbance explains the variations.

\begin{figure}[ht]
\centering
\begin{minipage}{0.50\linewidth}
\centering
\includegraphics[width=\linewidth]{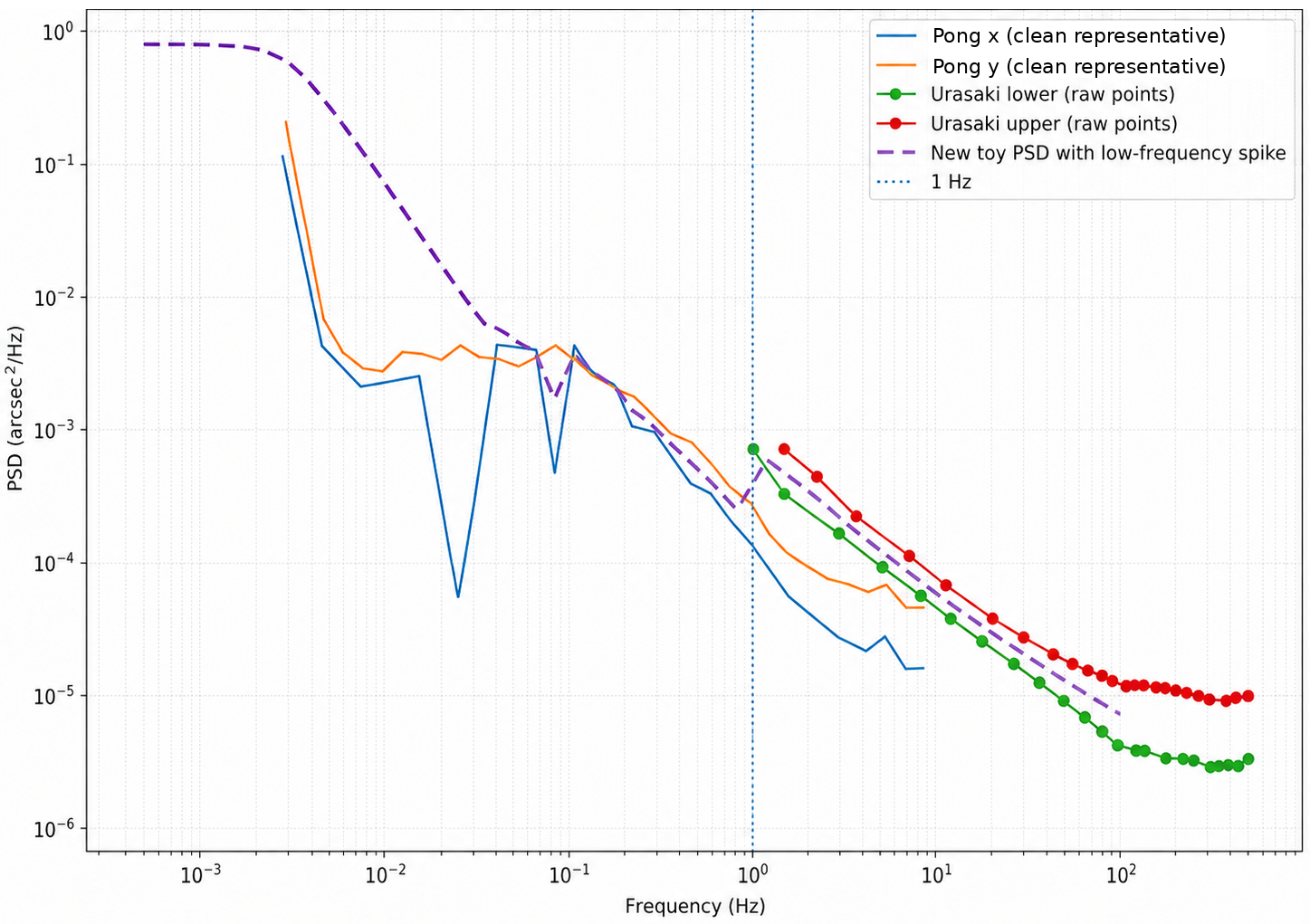}
\end{minipage}
\hfill
\begin{minipage}{0.38\linewidth}
\centering
\includegraphics[width=\linewidth]{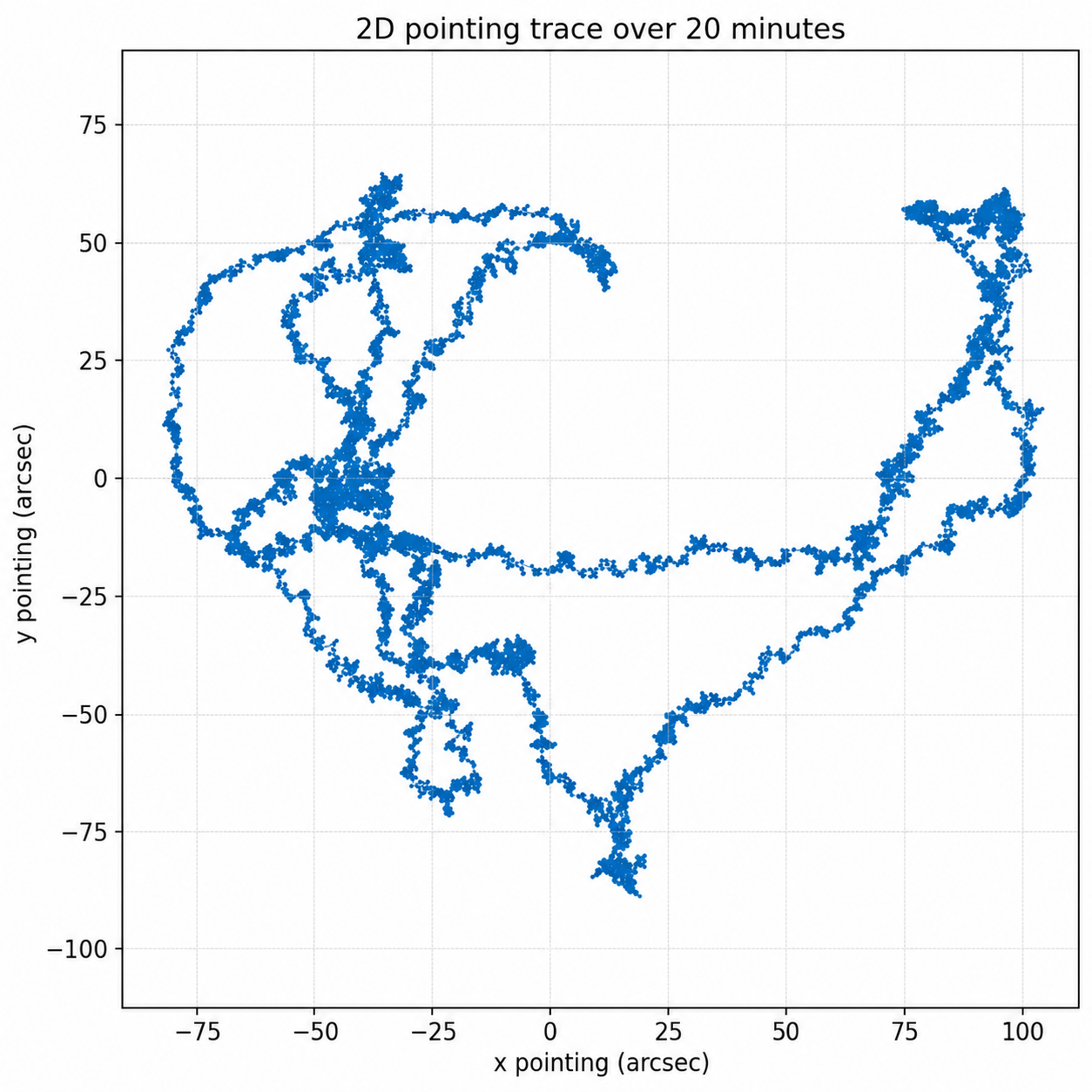}
\end{minipage}
\caption{Current hybrid pointing-error model used as an input disturbance
for PT simulations. Left: pointing-error power spectrum assembled from
resampled ASTERIA pointing behavior \cite{pong2018_asteria}, HyTI jitter
measurements \cite{urasaki2024_hyti}, and low-frequency content chosen to
reproduce the spacecraft-jitter allocation. Right: corresponding two-axis
pointing trajectory over a representative 20 minute simulation. The purpose
is not to claim a final bus jitter PSD, but to expose the PT/FID optical
loops to plausible low-frequency drift, reaction-wheel-like structure, and
camera-rate sampling before measured spacecraft data are available.}
\label{fig:tracking_jitter}
\end{figure}

\noindent{\bf Tracking-jitter model:} A key lesson from the SRR trade studies is that spacecraft pointing error cannot be treated as a single static number. For STARI, the relevant disturbance is a spectrum: low-frequency attitude motion can be tracked by the payload camera/FSM loop, while faster reaction-wheel or structural content appears as residual image motion and ultimately as fiber-coupling noise. Rather than using a single vendor stability value, the current simulation work builds a project-generated hybrid jitter power spectrum from CubeSat pointing and vibration data reported for ASTERIA and HyTI \cite{pong2018_asteria,urasaki2024_hyti}; see Figure\,\ref{fig:tracking_jitter} for representative calculation of the expected pointing jitter.

\begin{figure}[ht]
\centering
\includegraphics[width=0.86\linewidth]{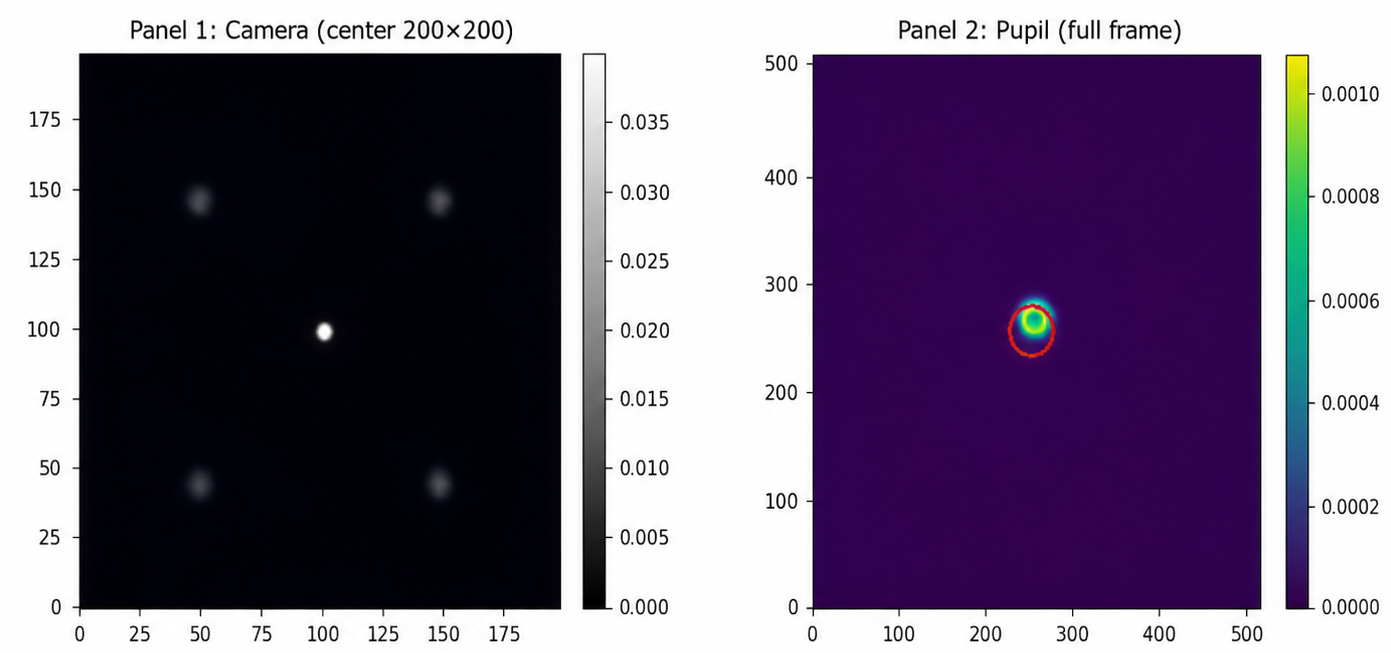}
\caption{Visualization from the current PT/FID pointing simulation.
 Panel 1 shows the image formed on the PT camera: the target star appears near the center while four LED beacons from the partner spacecraft define the relative spacecraft geometry. Panel 2 shows the resulting pupil/beam footprint (including Fresnel diffraction) projected onto the receiving aperture of the spacecraft acting as FID. }
\label{fig:pt_python_gui}
\end{figure}

\paragraph{Control-loop implications:}
The optical-control concept uses differential image information rather than
relying on absolute spacecraft attitude alone. The PT camera sees the target
star and the FID beacon pattern in the same frame, so common pointing errors
can be converted into an FSM command that keeps the outgoing pupil aligned
with the FID aperture. In the current simplified PT simulation, the input
disturbance is a synthesized two-axis spacecraft-jitter model with roughly
60 arcsec rms low-frequency motion and drifts below about 1 arcsec s$^{-1}$.
For the photon rates expected from a $V \simeq 2.5$ mag target, the PT
centroiding error is estimated to be $\sim 0.25$ arcsec in a 20 ms exposure.
Under these assumptions, a simple predictive/Kalman-style controller reduces
the residual one-axis tracking error to below 1 arcsec in representative
runs, comfortably below the 4 arcsec PT allocation for the baseline
100 m handoff case. At 100 m, a 1 arcsec residual corresponds to about
0.5 mm of pupil motion, compared with the $\sim 2$ mm pupil-centering
allocation used for the stable-fiber-coupling goal. Figure~\ref{fig:pt_python_gui} shows a representative PT camera view and the corresponding
received-pupil footprint used in the current simulation.

These results should be interpreted as an architecture-level consistency
check rather than a final performance prediction. The next model iteration
will include measured camera exposure/readout latency, centroid bias from
defocus and undersampling, optical transmission losses, beacon brightness,
FSM command delay and hysteresis, relative-orbit drift, and an improved model of the
spacecraft-bus jitter spectrum after bus selection.

\FloatBarrier

\section{Conclusion}
\label{sec:challenges}

STARI has evolved substantially since the 2024 SPIE update, moving from an
early mission concept toward a more mature integrated flight design for
demonstrating core technologies of future free-flying space interferometry.
The importance of STARI is not any single subsystem in isolation, but the
integration of beam handoff, formation control, propulsion, payload avionics,
and fiber injection into one flight experiment.
The present design has converged to a coupled optical, GNC, propulsion, and
spacecraft-bus architecture.
Near-term technology-maturation and validation work includes
final spacecraft-bus selection and
measured jitter spectra, camera exposure/readout latency, FSM command delay
and hysteresis, beacon brightness and stray-light behavior, optical
throughput, OAP/fiber thermal stability, focus-mechanism repeatability,
propulsion minimum-impulse characterization, and validation of autonomous
standby-to-science sequencing.

Near-term pacing items for the January 2027 Design Review include completing
bus/payload interface control, payload computer and camera/FSM procurement,
OAP and focus-mechanism trades, propulsion prototype testing, and a more
complete end-to-end software and autonomy architecture. Flight-unit
fabrication for the bus and subsystems is planned throughout 2027, followed
by system integration and environmental testing in 2028 and launch-readiness
activities for an expected 2029 launch.

\acknowledgments
STARI is supported by the NASA Astrophysics Research and Analysis (APRA) program. We thank the December 2025 SRR review committee and reviewers, including Eric Gustafson, David Hinkley, Rebecca Kelly, Fernando Saca, Dan Scharf, Terry Stevenson, Gerard van Belle, and Andy Klesh, for their detailed review and action items, which helped mature the requirements, risk posture, and path to PDR. We also thank our NASA colleagues John Hudeck, Luis Santos, and David Morris for guidance and support as the mission has matured. Lastly, we have appreciated ongoing discussions with colleagues Sylvestre Lacour, Michael Ireland, Jonah Hansen, Ito Takahiro, and Taro Matsuo. ChatGPT was used during the production of this manuscript. We recognize early seed support from the Michigan Institute for Research in Astrophysics.

\end{document}